\newcommand{\eqb}{\begin{equation}}
\newcommand{\eqe}{\end{equation}}
\documentclass[rmp, aps, amsfonts, amsmath, amssymb,  floatfix, superscriptaddress]{revtex4}

\usepackage{graphicx}
\usepackage{natbib}

\begin{document}

\title{Kinetics of protein-DNA interaction: 
facilitated target location in sequence-dependent potential}

\author{Michael Slutsky}
\affiliation{Department of Physics, Massachusetts Institute of
Technology, 77 Massachusetts Avenue, Cambridge, MA 02139, USA}

\author{Leonid A. Mirny}
\affiliation{Department of Physics, Massachusetts Institute of
Technology, 77 Massachusetts Avenue, Cambridge, MA 02139, USA}
\affiliation{Harvard-MIT Division of Health
Sciences and Technology, Massachusetts Institute of Technology, 
77 Massachusetts Avenue, Cambridge, MA 02139, USA}

\begin{abstract}
Recognition and binding of specific sites on DNA by proteins is
central for many cellular functions such as transcription,
replication, and recombination. In the process of recognition, a
protein rapidly searches for its specific site on a long DNA
molecule and then strongly binds this site. Here we aim to find a
mechanism that can provide both a fast search (1-10 sec) and high
stability of the specific protein-DNA complex
($K_d=10^{-15}-10^{-8}$ M).

Earlier studies have suggested that rapid search involves the sliding
of a protein along the DNA. Here we consider sliding as a
one-dimensional (1D) diffusion in a sequence-dependent rough energy
landscape. We demonstrate that, in spite of the landscape's roughness,
rapid search can be achieved if 1D sliding is accompanied by 3D
diffusion. We estimate the range of the specific and non-specific
DNA-binding energy required for rapid search and suggest experiments
that can test our mechanism. We show that optimal search requires a
protein to spend half of time sliding along the DNA and half diffusing
in 3D. We also establish that, paradoxically, realistic energy
functions cannot provide both rapid search and strong binding of a
rigid protein. To reconcile these two fundamental requirements we
propose a search-and-fold mechanism that involves the coupling of
protein binding and partial protein folding.
				   
Proposed mechanism has several important biological implications for
search in the presence of other proteins and nucleosomes, simultaneous
search by several proteins etc. Proposed mechanism also provides a new
framework for interpretation of experimental and structural data on
protein-DNA interactions.
\end{abstract}

\maketitle

\section{Introduction}
The complex transcription machinery of cells is primarily regulated by
a set of proteins, {\sl transcription factors} (TFs), that bind DNA at
specific sites. Every TF can have from one to several dozens of such
specific sites on the DNA.  Upon binding to the site, TF forms a
stable protein-DNA complex that can either activate or repress
transcription of nearby genes, depending on the actual control
mechanism. Fast and reliable regulation of gene expression requires
(1) fast ($\sim$1-10 sec) search and recognition of the specific site
(referred to as the {\em target } or {\em cognate} site below) out of
$10^6$ - $10^9$ possible sites on the DNA, and (2) stability of the
protein-DNA complex ($K_d=10^{-15}-10^{-8}$ M). In spite of its
apparent simplicity, such a mechanism is not understood in depth,
either qualitatively or quantitatively. Here we focus on a simpler
case of bacterial TFs recognizing their cognate (target) sites on the
naked DNA.  Needless to say that eukaryotic protein-DNA recognition is
significantly complicated by chromatin packing of the DNA and
multi-subunit structure of TFs. Interestingly, similar problems of
specific binding and binding rate arises in the context of
oligonucleotides-DNA binding \cite{lomakin98}

Vast amounts of experimental data available these days provide the
structures of protein-DNA complexes at atomic resolution in crystals
and in solution
\cite{luscombe,lewis_bell1,lewis_bell2,lewis,schumacher}, binding
constants for dozens of native and hundreds of mutated proteins
\cite{sarai_pnas,trp}, calorimetry measurements
\cite{spolar}, and novel single-molecule experiments
\cite{shimamoto}. These experimental data contributed most
significantly to our present understanding of protein-DNA interaction
since the early work of von Hippel, Berg $et.~al.$ In a series of
pioneering articles \cite{bvh_kin1,bvh_kin3,bvh_kin2,bvh_therm}, they
have created a conceptual basis for describing of both the kinetics
and thermodynamics of protein-DNA interaction, which became a starting
point for practically every subsequent theoretical work on the
subject.

We start by reviewing the history of the problem and describing the
paradox of the "faster than diffusion" association rate. Next, we
present the classical model of protein-DNA "sliding" and explain how
this model can resolve the paradox. We outline the problem that the
sliding mechanism faces if the energetics of protein-DNA interactions
are taken into account. Next we introduce our novel quantitative
formalism and undertake in-depth exploration of possible mechanisms of
protein-DNA interaction.

\subsection{"Faster than diffusion" search} 
The problem of how a protein finds its target site on DNA has a
long history. In 1970, Riggs et.~al. \cite{riggs1,riggs2} measured
the association rate of LacI repressor and its operator on DNA as
$ \sim 10^{10}$ M$^{-1}$s$^{-1}$. This astonishingly high rate (as
compared to other biological binding rates) was shown to be much
higher than the maximal rate achievable by 3D diffusion. In fact,
if a protein binds its site by 3D diffusion, it has to hit the
right site on the DNA within $b=0.34~$nm. (A shift by $0.34~$nm
would result in binding a site that is different from the native
one by 1bp. Such a site can be very different, e.g. GCGCAATT vs
CGCAATTC).  Using the Debye-Smoluchowski equation for the {\sl
maximal} rate of a bimolecular reaction (see e.~g.
\cite{eigen,bruinsma,bruinsma1}), with a protein diffusion
coefficient of $D_{3d}\sim 10^{-7}$cm$^2/$s \cite{elowitz} we get
\begin{equation}
k_{DS} = 4\pi D_{3D} b \sim 10^8 ~\mbox{M}^{-1} \mbox{s}^{-1}
\end{equation}
This value for the association rate, relevant for {\it in vitro}
measurements, corresponds to target location {\it in vivo} on a
time scale of a few seconds, when each cell contains up to several
tens of TF molecules.

To resolve the discrepancy between the experimentally measured rate of
$10^{10}$ M$^{-1}$s$^{-1}$ and the maximal rate of $10^{8}$
M$^{-1}$s$^{-1}$ allowed by diffusion, Riggs et.~al., Richter
et.~al. \cite{eigen} and later Winter, Berg and von Hippel
\cite{bvh_kin1,bvh_kin2} suggested that the dimensionality of the
problem changes during the search process. They concluded that while
searching for its target site, the protein periodically scans the DNA
by ``sliding'' along it.

\subsection{Sliding along the DNA} 
If a protein performs both 3D and 1D diffusion, then the total search
process can be considered as a 3D search followed by binding DNA and a
round of 1D diffusion. Upon dissociation from the DNA, the protein
continues 3D diffusion until it binds DNA in a different place, and so
on. Some experimental evidence supports this search mechanism. These
include affinity of the DNA-binding proteins for any fragment of DNA
(non-specific binding), single molecule experiments where 1D diffusion
has been observed and visualized, and numerous other experiments where
the rate of specific binding to the target site has been significantly
increased by lengthening non-specific DNA surrounding the site
\cite{kim}.  What are the benefits and the mechanism of 1D diffusion
and what limits the search rate?

Here we address this question and consider possible search mechanisms
that involve both 1D and 3D diffusion, where 1D diffusion along
the DNA proceeds along the rough energy landscape.  Quantitative
analysis of the search process brought us to the following four main
results:
\begin{itemize}
\item When the roughness of the binding energy landscape is
greater than ${\sim 2 k_BT}$, the diffusion along the DNA becomes
extremely slow with the protein unable to diffuse more than a
few base-pairs. The total search process is prohibitively slow.

\item If the search proceeds by a combination of 1D and 3D
diffusion, non-specific binding to the DNA plays a very important
role in controlling the balance between these two processes. The
optimal energy of non-specific binding can provide the maximal
search rate. Although faster than either 3D or 1D search alone,
optimal combination of 3D and 1D diffusion cannot expedite the search
if the roughness of the landscape is greater than $\sim 2
k_BT$.

\item Experimentally observed and biologically relevant rates of
search can be reached only when 1D sliding proceeds through a fairly
smooth landscape with a roughness of the order of $k_BT$.

\item Paradoxically, the stability of the protein-DNA complex at the
target site requires a roughness of the binding energy landscape
considerably larger than $k_BT$. Rapid search, however, by 1D/3D
diffusion is impossible at such roughness.
\end{itemize}

Finally, we formulate this ``search speed--stability'' paradox and
suggest a search-and-fold mechanism that can resolve it.  The paradox
can be resolved if the DNA-binding protein has two distinct
(conformational) states in which it exhibits two modes of binding. In
the first, mode that has weaker binding and a smother landscape, it
searches for its site. In the second recognition mode, that has larger
roughness of the binding landscape, the protein tightly binds DNA
sites. Correlation between the energy landscapes in the two modes and
the energy difference and the barrier between the two protein
conformations control the frequency of transition between the two
modes and provides effective pre-selection of low-energy sites.

We suggest that these modes correspond to two distinct conformational
states of the protein-DNA complex (a more open complex in the search
mode, and a more tight one in the recognition mode). Transition
between the two states can include partial folding of the protein,
water extrusion, change in the DNA conformation etc. Focusing on the
conformation of the protein, and without loss of generality, we
consider a partially unfolded (disordered) conformation and the folded
conformation bound to the cognate site as the two conformations
required by our model. In fact, a protein in the partially unfolded
conformation may have fewer and/or weaker interactions with DNA
allowing rapid sliding. Folded conformation, in turn, provides
stronger and more specific interactions required for for tight
binding.

We also quantify the requirements of this two-mode mechanism to
provide {\em both} rapid search and stability. Structures of known
DNA-binding protein are known to be flexible and have been reported to
exhibit two or more distinct binding modes. This two-state mechanism
also agrees well with the results of calorimetric experiments.

Proposed search-and-fold mechanism is not limited to protein-DNA
interaction providing a general framework for protein-ligand
binding and demonstrating advantages of induced folding, a common
theme in molecular recognition.

\section{The Model} 
\subsection{Search time} 
In our model, the search process consists of $N$ rounds of 1D search
(each takes time of $\tau_{1d,i}, i=1..N$) separated by rounds of 3D
diffusion ($\tau_{3d,i}$). The total search time $t_s$ is the sum of
the times of individual search rounds:
\begin{equation}
\label{eq:start}
t_s = \sum_{i = 1}^N\left(\tau_{1d,i}+\tau_{3d,i}\right).
\end{equation}
The total number $N$ of such rounds occurring before the target
site is eventually found is very large, so it is natural to
introduce probability distributions for the essentially random
entities in the problem.  The first obvious simplification that
can be made without any loss of rigor is to replace $\tau_{3d,i}$
by its average $\tau_{3d}$. Each round of 1D diffusion scans a
region of $n$ sites (where $n$ is drawn from some distribution
$p(n)$). The time $\tau_{1d}(n)$ it takes to scan $n$ sites can be
obtained from the exact form of the 1D diffusion law (see Appendix
A). If, on average, ${\bar n}$ sites are scanned in each round,
then the average number of such rounds required to find the site
on DNA of length $M$ is $N=M/{\bar n}$. Using average values,
we get a total search time of
\begin{equation}
\label{eq:main} t_s \left( \bar{n} , M \right) =
\frac{M}{{\bar n}} \left[\tau_{1d}\left( \bar{n} \right) +
\bar{\tau}_{3d}\right],
\end{equation}
From (\ref{eq:main}) it is clear that in general, $t_s\left(
\bar{n} , M \right)$ is large for both very small and very large
values of $\bar{n}$. In fact if $\bar n$ is small, very few sites are
scanned in each round of 1D search and a large number of such rounds
(alternating with rounds of 3D diffusion) are required to find the
site. On the contrary, if $\bar n$ is large, lots of time is spent
scanning a single stretch of DNA, making the search very redundant and
inefficient. An optimal value $\bar{n}_{\rm opt}$ should exist that
provides little redundancy of 1D diffusion and a sufficiently small
number of such rounds. For a given diffusion law $\tau_{1d}(n)$,
function $t_s\left( \bar{n} , M \right)$ can be minimized producing
$\bar{n}_{\rm opt}$, the optimal length of DNA to be scanned between
the association and the dissociation events \footnote{Naturally, we
assume here that $\tau_{1d}\left(
\bar{n} \right)$ grows with  $\bar{n}$ at least as
$O(\bar{n}^{1+\alpha})$, with $\alpha > 0$.}.

\subsection{Protein-DNA energetics} 
While diffusing along DNA, a TF experiences the binding
potential $U({\vec s})$ of every site $\vec s$ it encounters. The
energy of protein-DNA interactions is usually divided into two
parts, {\it specific} and {\it non-specific} \cite{bvh_therm,hwa}
\eqb U_i = U(\vec{s}=s_i,..s_{i+l-1}) + E_{\rm ns}, \eqe where
$\vec{s}$ describes a binding DNA sequence of length $l$. As its name
suggests, the non-specific binding energy $E_{\rm ns}$ arises from
interactions that do not depend on the DNA sequence that the TF is
bound to, e.~g. interactions with the phosphate backbone.  The
specific part of the interaction energy exhibits a very strong
dependence on the actual nucleotide sequence. Here and below we use
the term "energy" referred to the change in the free energy related to
binding $\Delta G_{b}$. This free energy includes the entropic loss of
translational and rotational degrees of freedom of the protein and
amino acids' side-chains, the entropic cost of water and ion extrusion
from the DNA interface, the hydrophobic effect etc.

\begin{figure*}[h]
\begin{center}
\includegraphics[width = 0.7\textwidth]{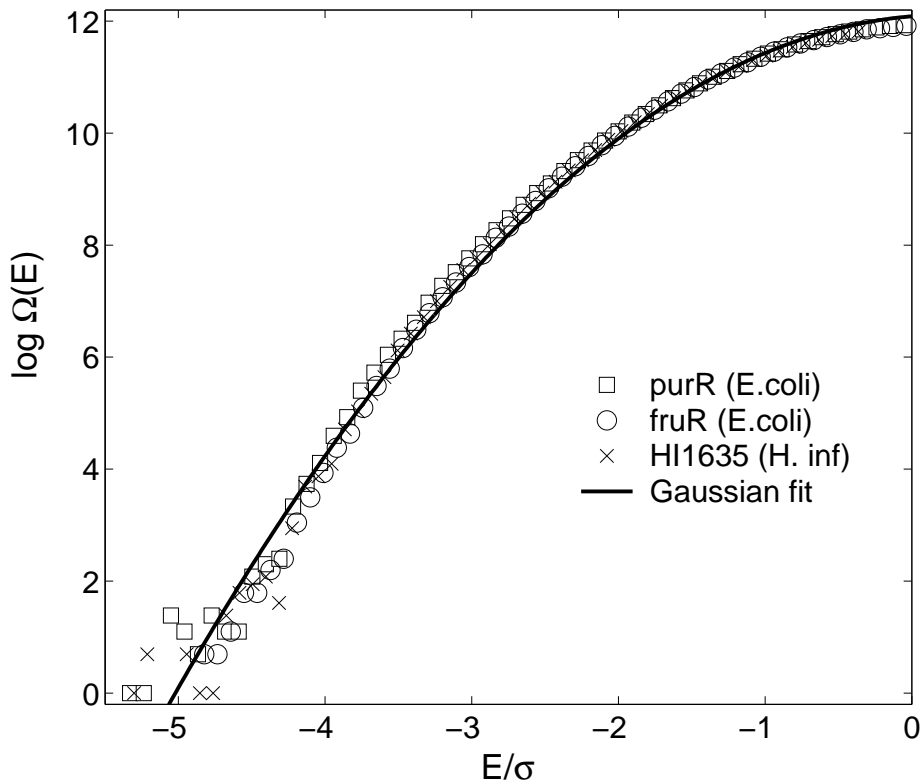}
\end{center}
\caption{Spectrum of binding energy for three different
transcription factors and the Gaussian approximation (solid line).} \label{fig:no_gap}
\end{figure*}

The energy of specific protein-DNA interactions can be
approximated by a weight matrix (also known as "PSSM" or
"profile") where each nucleotide contributes independently to the
binding energy \cite{bvh_therm}:
\begin{equation} \label{eq:U}
U(\vec{s}=s_i,..s_{i+l-1}) = \sum_{j=1}^{l} \epsilon(j,s_j),
\end{equation}
where $s_j$ is a base-pair in position $j$ of the site and
$\epsilon(j,x)$ is the contribution of base-pair $x$ in position
$j$. Most of the known weight matrices of TFs $\epsilon(j,s_j)$ give
rise to uncorrelated energies of overlapping neighboring sites,
obtained by one base pair shift \cite{hwa}. Figure~\ref{fig:no_gap}
presents distributions of the sequence specific binding energy
$f(U)$ obtained for different bacterial transcription factors and
all possible sites in the corresponding genome. The weight
matrices for these transcription factors has been derived using a
set of known binding sites and standard approximation
\cite{bvh_therm,stormo_fields}. Notice that for a sufficiently long
site the distribution of the binding energy of random sites (or
genomic DNA) can be closely approximated (see Fig.~\ref{fig:no_gap})
by a Gaussian distribution with a certain mean $\langle U\rangle$
and variance $\sigma^2$:
\begin{equation}
f(U_i) = \frac{1}{\sqrt{2\pi\sigma^2}} \exp\left[-\frac{\left(U_i
- \langle U\rangle\right)^2}{2\sigma^2}\right].
\end{equation}
We also assume independence of the energy of neighboring (though
overlapping) sites. Binding energies calculated for bacterial TFs
support this assumption. Other physical factors such as local DNA
flexibility \cite{bustamante_cro} can create a correlated energy
landscape providing a different mode of diffusion that we have
described in \cite{skm_condmat}.

\subsection{Diffusion in a sequence-dependent energy landscape} \label{sec:diffusion} 
\begin{figure*}[h]
\begin{center}
\includegraphics[width = 0.7\textwidth]{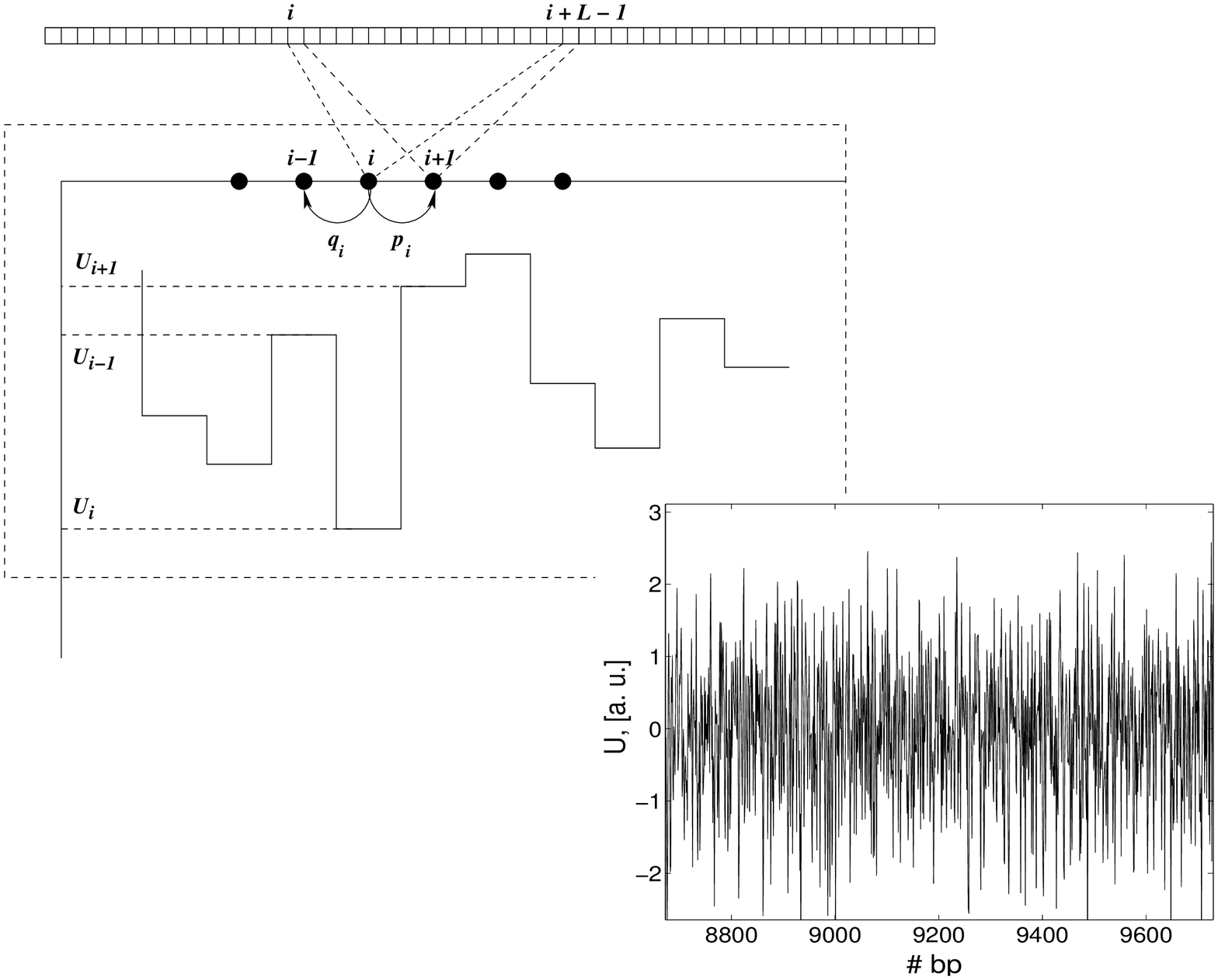}
\end{center}
\caption{ The Model Potential.} \label{fig:potential}
\end{figure*}

The whole DNA molecule can thus be mapped onto one-dimensional
array of sites $\left\{\vec{s}_i\right\}$, each corresponding to
a certain binding sequence comprising bases from the $i$-th to
the $(i+l-1)$-th, $l$ being the length of the motif (see
Fig.~\ref{fig:potential}). At each site, there is a probability
$p_i$ of hopping to site ${i+1}$ and a probability $q_i$ of
hopping to site ${i-1}$. These probabilities depend on the
specific binding energies $U_i$ and $U_{i \pm 1}$ at the $i$-th
site and at the adjacent sites respectively and are proportional
to the corresponding transition {\em rates,} $\omega_{i,i+1}$ and
$\omega_{i,i-1}$. For the latter, it is most natural to assume
the regular activated transport form

\begin{equation}\label{eq:prob} 
\omega_{i,i\pm 1} = \nu \times  \left\{
\begin{array}{l}
e^{ -\beta\left(U_{i\pm 1} -U_{i}\right)} \qquad {\rm if}~U_{i\pm 1} > U_{i} \\
1.0 \qquad\qquad {\rm otherwise }
\end{array}
\right.,  
\end{equation} 
where $\nu$ is the effective attempt frequency, $\beta\equiv\left(k_BT\right)^{-1}$, $k_B$ is the Boltzmann constant and $T$ is
the ambient temperature. Having defined that, we have a
one-dimensional random walk with position-dependent hopping
probabilities.

As has been shown in numerous papers throughout the last two
decades, the properties of 1D random walks can vary dramatically
depending on the actual choice of probabilities $\left\{
p_i\right\}$ (for review, see e.~g. \cite{bouchaud}). Here we
employ the mean first-passage time formalism \cite{kehr} to derive
the diffusion law $\tau_{1d}\left( \bar{n} \right)$ for protein
sliding along the DNA given the sequence-dependent binding energy
(\ref{eq:prob}).

\section{Results} 
Using the model described above, we studied the following
problems:
\begin{itemize}
\item How fast is the 1D search on DNA as a function of the
``roughness'' $\sigma$ of the binding energy landscape?
\item How significant is the role of non-specific binding energy $E_{\rm ns}$ in
determining the search time?
\item How fast is the search for the native site under conditions that
provide stability to the protein-DNA complex at the target site?
\end{itemize}

\subsection*{Diffusion along the DNA} 
We state here the main results without a derivation (which can be
found in the Appendix A). For a given set of probabilities
$\left\{ p_i\right\}$, the mean first-passage time (MFPT) from $i
= 0$ to $i = L$ (in terms of number of steps) is \cite{kehr}
\begin{equation}\label{eq:mfpt}
\bar{t}_{0,L} = L + \sum _{k = 0}^{L - 1} \alpha_k + \sum _{k =
0}^{L - 2} \sum _{i = k + 1}^{L - 1} \left( 1 + \alpha_k \right)
\prod_{j = k + 1}^{i} \alpha_j, \end{equation} where $\alpha_i \equiv
q_i/p_i$.  The relation (\ref{eq:mfpt}) gives the MFPT for one given
realization of probabilities. Assuming that the specific binding
energies $\left\{ U_i\right\}$ have a normal distribution with
variance $\sigma^2$ (see above), we plug the probabilities in
(\ref{eq:prob}) into (\ref{eq:mfpt}) and after a somewhat lengthy but
straightforward calculation, we obtain an expression for the MFPT
averaged over genomic sequences for $L \gg 1$:

\begin{equation}\label{eq:mfpt_fin}
\left\langle\bar{t}_{FP}\left(L\right)\right\rangle \simeq \tau_0
L^2 e ^{7\beta^2\sigma^2/4}\left(1 + \beta^2\sigma^2/2
\right)^{-1/2},
\end{equation}
where $\tau_0$ is the reciprocal of the effective attempt
frequency for hopping to a neighboring site.

The main result is that the 1D search by hopping to neighboring
sites proceeds by normal diffusion with $t \sim L^2/2D_{1d}$, where
the diffusion coefficient
\begin{equation}
D_{1d}\left(\sigma\right) \simeq \frac{1}{2\tau_0}\left(1 +
    \frac{\beta^2\sigma^2}{2}\right)^{1/2} e^{- 7\beta^2\sigma^2/4}
\end{equation}
exhibits an exponential dependence on the ``roughness'' of the
binding energy landscape $\sigma$, dropping rapidly as $\sigma$
becomes greater than few $k_BT$ \cite{skm_condmat}. Hence, rapid
diffusion of a protein along the DNA is possible only if the roughness
of the binding energy landscape is small compared to $k_BT$
($\beta\sigma < 1.5$). This requirement imposes strong constraints
on the allowed energy of specific binding interactions.

\subsection{Optimal time of 3D/1D search} 
When 1D scanning is combined with 3D diffusion, what is the
optimal time a protein has to spend in each of the two regimes? To
answer this question we compute the optimal number of sites the
protein has to scan by 1D diffusion in order to get the fastest
overall search. Results of this section are rather general and are
not limited to the particular scenario of slow 1D diffusion on a
rough landscape discussed above.

Each time the protein binds DNA it performs a round of 1D
diffusion. If the round lasts $\tau_{1d}$ then on average the
protein scans $\bar{n}= \sqrt{16 D_{1d} \tau_{1d} / \pi}$
bps. \cite{hughes} By plugging this relation into
Eq.~(\ref{eq:main}) for search time $t_s$, and minimizing $t_s$
with respect to $\bar{n}$, we get the optimal total search time and
the optimal number of sites to be scanned in each round:

\begin{equation}\label{eq:ts_fin_min}
t_s^{\rm opt}=t_s(\bar{n}_{\rm opt}) = \frac{M}{2}\sqrt{\frac{\pi
\bar{\tau}_{3d}}{D_{1d}}} \ \ \ \ \bar{n}_{\rm opt} =
\sqrt{\frac{16}{\pi} D_{1d}\bar{\tau}_{3d} }
\end{equation}

This analysis brings us to the following conclusions.

First, and most importantly, we obtain that in the \emph{optimal}
regime of search
\begin{equation}
\tau_{1d} (\bar{n}_{\rm opt}) = \tau_{3d},
\end{equation}
i.e. the protein spends equal amounts of time diffusing along
non-specific DNA and diffusing in the solution. This striking
result is very general, and is true irrespective of the values of diffusion
coefficients $D_{1d}$ or $D_{3d}$, or size of the genome $M$. In
fact it follows directly from the diffusion law $\bar{n} \sim
\sqrt{\tau_{1d}}$. More importantly this central result can be
verified experimentally by either single-molecule techniques or by
traditional methods.

Also note that the optimal region of the DNA scanned in a single
round of 1D diffusion $\bar{n}_{\rm opt}$ does not depend on M,
i.e. is the same irrespective of the size of the genomes to be
searched for a specific site.

Second, the optimal 1D/3D combination reached at $\tau_{1d}=\tau_{3d}$
leads to a significant speed up of the search process. In fact, an
optimal 1D/3D search is $\bar{n}_{\rm opt}$ times faster than a search
by 3D diffusion alone, and ${M}/{\bar{n}_{\rm opt}}$ times faster
than a search by 1D diffusion alone. For example, if the protein
operates in the optimal 1D/3D regime and scans $\bar{n}_{\rm
opt}=100$bp during each round of DNA binding, then the experimentally
measured rate of binding to the specific site can be $100$ times
greater than the rate achievable by 3D diffusion alone.

Third, we can estimate  $\bar{n}_{opt}$, the maximal number of sites a protein can
scan in each round of 1D search. If we set
$D_{1d}$ to its maximum, i.~e. $D_{1d}\sim D_{3d}$ and
$\bar{\tau}_{3d}\sim l_{\rm d}^2/D_{3d}$, with $l_{\rm m}\sim 0.1
\mu{\rm m}$, we get
\begin{equation}
\bar{n}_{\rm opt}^{\rm max}\sim 500 ~{\rm bp}.
\label{eq:nopt}
\end{equation}
For a smaller 1D diffusion coefficient, e.~g. $D_{1d}\sim D_{3d}
/100$, we get  $\bar{n}_{\rm opt}^{\rm max}\sim 50$bp. Again,
single molecule experiments can provide estimates of these
quantities for different conditions of diffusion.

Finally, we obtain estimates of the shortest possible total search
time. If $M\approx 10^6$~bp and 1D diffusion is at its fastest
rate, i.~e. $D_{1d}\sim D_{3d}= 10^{-7}$cm${^2}/$s, then using
Eq.~(\ref{eq:ts_fin_min}) we get
\begin{equation}\label{eq:s_time}
t^{opt}_{s} \sim \frac{M}{2}\sqrt{2\pi \bar{\tau}_{3d}\tau_0}\sim
5~{\rm sec},
\end{equation}
where we estimate $\tau_0\sim a_0^2/D_{3d}\sim 10^{-8}$ sec.

One can also estimate the search time using {\it in vitro}
experimentally measured binding rates in water $k^{\rm water}_{on}
\approx 10^{10}$M$^{-1}$s$^{-1}$ \cite{riggs1,riggs2}. The
diffusion coefficient of a protein in the cytoplasm is $10-100$
times lower than that in water leading to the estimated binding
rate of $k^{\rm cytoplasm}_{on} \approx
10^{8}-10^{9}$M$^{-1}$s$^{-1}$ (see Appendix D). From this we
obtain the time it takes for one protein to bind one site in a
cell of $1\mu{\rm m}^3$ volume (i.e. [TF]$\approx 10^{-9}$M) as
\begin{equation}
t^{exp}_{s}={\left(k^{\rm cytoplasm}_{on} {\rm [TF]}\right)^{-1}}
\sim 1-10~{ \rm sec}.
\label{eq:tsearchTF}
\end{equation}
One can see perfect agreement between our theoretical estimates
and experimentally measured binding rates.

As we mentioned above, there are usually several TF molecules
searching in parallel for the target site. Naturally, in this
case, the search is sped up proportionally to the number of
molecules.

\subsection{Diffusion of PurR on E.coli genome} 
To check the applicability of the above considerations, we simulated
one - dimensional diffusion of $PurR$ transcription factor on the
$E.~coli$ chromosome.

\begin{figure*}[h]
\begin{tabular}{cc}
\includegraphics[width = 0.7\textwidth]{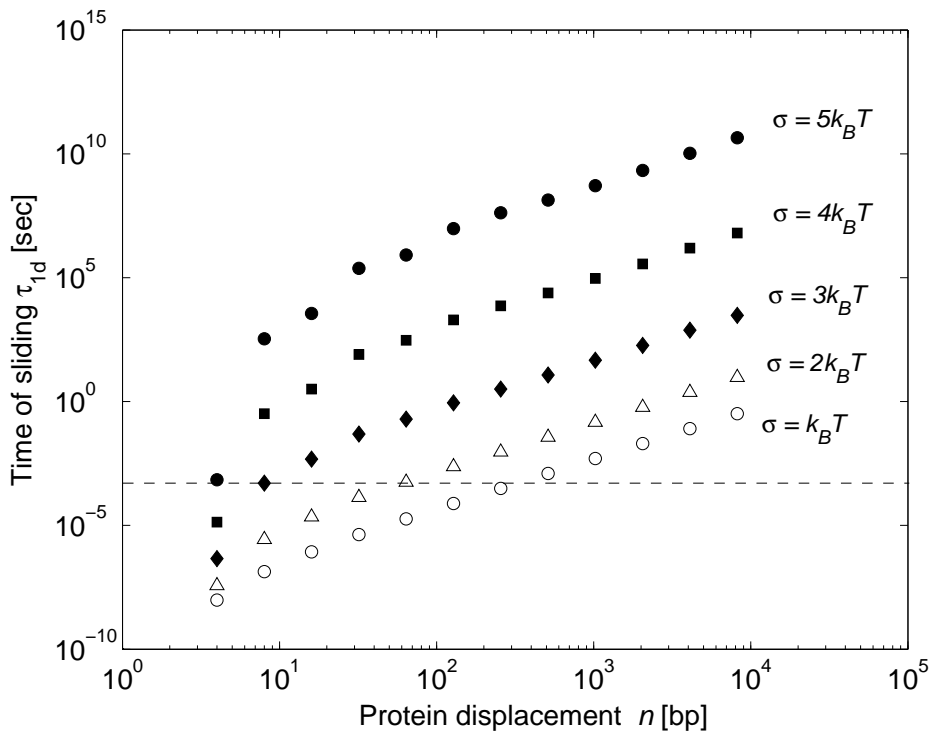}&
\end{tabular}
\caption{ The mean first passage time vs traveling distance for
$purR$ transcription factor on the binding landscapes of different
roughness (or at different temperatures). The horizontal line
indicates the optimal regime, $\tau_{1d} \sim \bar\tau_{3d}$.} \label{fig:pur_times}
\end{figure*}

The specific energy profile was built using a weight matrix
derived from 35 $PurR$ binding sites following a standard
procedure described elsewhere \cite{bvh_therm},
\cite{stormo_fields}. The resulting energy profile is random and
uncorrelated and has a standard deviation $\sigma \simeq
6.5~k_BT$. This profile was used as an input for calculating mean
first passage time at different temperatures\footnote{Since the
magnitude of the interaction is fixed, in these calculations we
vary temperature rather than binding strength}. The result of
these calculations is presented in Fig.~\ref{fig:pur_times}. It is
clear that when the roughness of the landscape becomes significant
$ \sigma  > 2 k_BT$, the diffusion proceeds extremely slowly.
Only $\sim 10-100$ bp can be scanned by a TF when $ \sigma=2
k_BT$. A natural requirement for sufficiently fast diffusion is,
as before, $\sigma \sim k_BT$.

\subsection{Non-specific binding} 
While the diffusion of the TF molecules along DNA is controlled by the
specific binding energy, the dissociation of the TF from the DNA
depends on the total binding energy, i.~e.  on the non-specific
binding as well as on the specific one.  Moreover, since the
dissociation events are much less frequent than the hopping between
neighboring base-pairs (roughly by a factor of
$\bar{\tau}_{3d}/\left\langle \tau \right\rangle$), the non-specific
energy $E_{\rm ns}$ makes a sensibly larger contribution to the total
binding energy.

For a TF at rest  bound to some DNA  site  $i$, the dissociation  rate
$r_i$ would be given by the Arrhenius - type relation,
\begin{equation}
r_i = \frac{1}{\tau_0} e^{ -\beta\left(E_{\rm ns} - U_i\right)}.
\end{equation}
Given the specific $U_i$ non-specific $E_{\rm ns}$ energy one can
calculate the average time $\tau_{1d}$ a protein spends before
dissociating from the DNA (see Appendix B). We obtain

\begin{equation}\label{eq:E_ns_fin} E_{\rm ns} =
k_BT\left[\ln\left(\frac{\tau_{1d}}{\tau_0}\right) -
\frac{1}{2}\left(\frac{\sigma}{k_BT}\right)^2\right],
\end{equation}
and in the optimal regime where $\tau_{1d} = \bar{\tau}_{3d}$
\begin{equation}\label{eq:E_ns_opt}
E^{\rm opt}_{\rm ns} =
k_BT\left[\ln\left(\frac{\tau_{3d}}{\tau_0}\right) -
\frac{1}{2}\left(\frac{\sigma}{k_BT}\right)^2\right].
\end{equation}

\subsection{The parameter space} 
\begin{figure*}[h]
\begin{tabular}{cc}
\includegraphics[width = 0.7\textwidth]{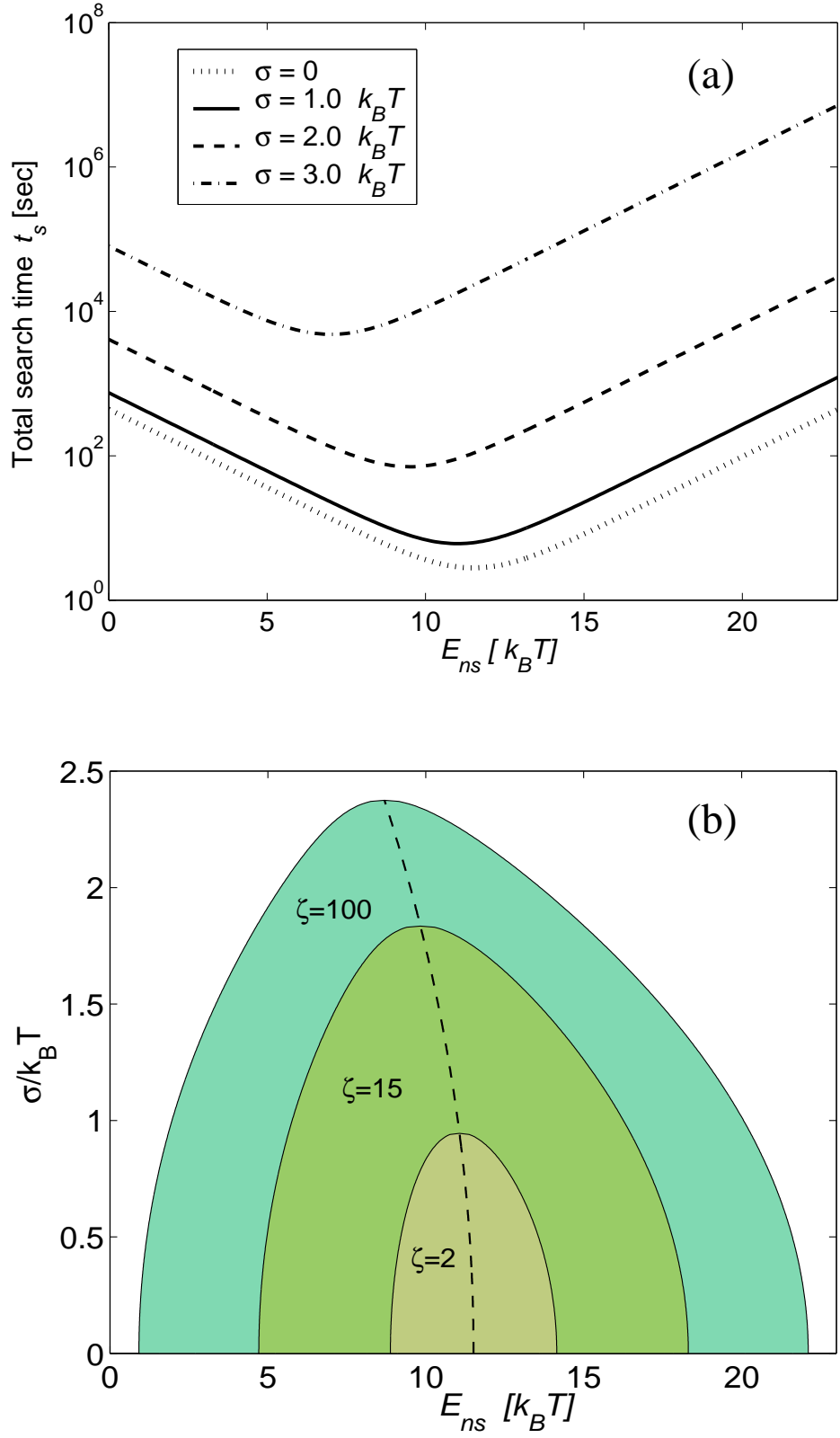}&
\end{tabular}
\caption{(a) Dependence of the search time on the
non-specific binding energy.  (b) The parameter space. The dashed
line corresponds to optimal parameters $\sigma$ and $E_{ns}$
connected by Eq. (\ref{eq:E_ns_opt}).} \label{fig:e_ns}
\end{figure*}

Since for a given value of $\sigma$, the non-specific binding
controls the dissociation rate, the search time will deviate from
the optimum if $E_{\rm ns}$ moves from this predetermined value.
In Fig.~\ref{fig:e_ns}a we plot the search time as a function of
the non-specific binding energy for different values of $\sigma$.

We now define the {\sl tolerance factor} $\zeta$ as the ratio
between the acceptable value of the search time $t_s$ and the
optimal search time $t^{\rm opt}_{s}$. Experimental data suggest
$\zeta \leq 5$, but for the moment we allow for much larger values
of $\zeta\sim 10 - 100$ (this can be done when, for instance,
there are many protein molecules searching in parallel). As we can
see from Fig.~\ref{fig:e_ns}a, for each value of $\sigma$, there
is a range of possible values of $E_{\rm ns}$ such that the
resulting search time is within the region of tolerance (see
Appendix B). Notice the dramatic increase in the search time as
$E_{\rm ns}$ deviates from its optimal value.

Specifying $\zeta$, we can define  our parameter space, i.~e. the
values of specific and non-specific energy producing a total search
time within the region of tolerance. In Fig.~\ref{fig:e_ns}b, we
consider three values of  $\zeta$. The most relaxed requirement
$\zeta=100$ provides a search time $t_{s} \leq 500~{\rm sec}$. If
$100$ proteins are searching for a single site, then the first one
will find it after $ \sim 5~{\rm sec}$, leading however to a
fairly low binding rate of $k_{\rm on} \approx 1/500 ~\rm{sec}
\cdot 10^{9} ~{\rm M}^{-1} = 2 \cdot 10^{6} ~{\rm M}^{-1} {\rm
s}^{-1}$ (compared to experimentally measured $10^{10} ~{\rm M}^{-1}
{\rm s}^{-1}$ in water). Importantly, in order to comply with even this most
relaxed search time requirement, the characteristic strength of
specific interaction must be smaller than $\sim 2.3 ~k_BT$.

These results bring us to a very important conclusion that a
protein cannot find its site in biologically relevant time if the
roughness of the specific binding landscape is greater than $\sim
2 ~k_BT$. Although an optimal 1D/3D combination can speed up the
search, it cannot overcome the slowdown of 1D diffusion. Only
fairly smooth landscapes ($\sigma \sim 1 k_BT$) can be effectively
navigated by proteins.

\section{Speed versus stability}
While rapid search requires fairly smooth landscapes ($\sigma \sim 1
k_BT$), stability of the protein-DNA complex, in turn, requires a
low energy of the target site ($U_{min} < 15~k_BT$ for a genome of $10^6$ bp).

\begin{figure*}[h]
\begin{tabular}{cc}
\includegraphics[width = 0.7\textwidth]{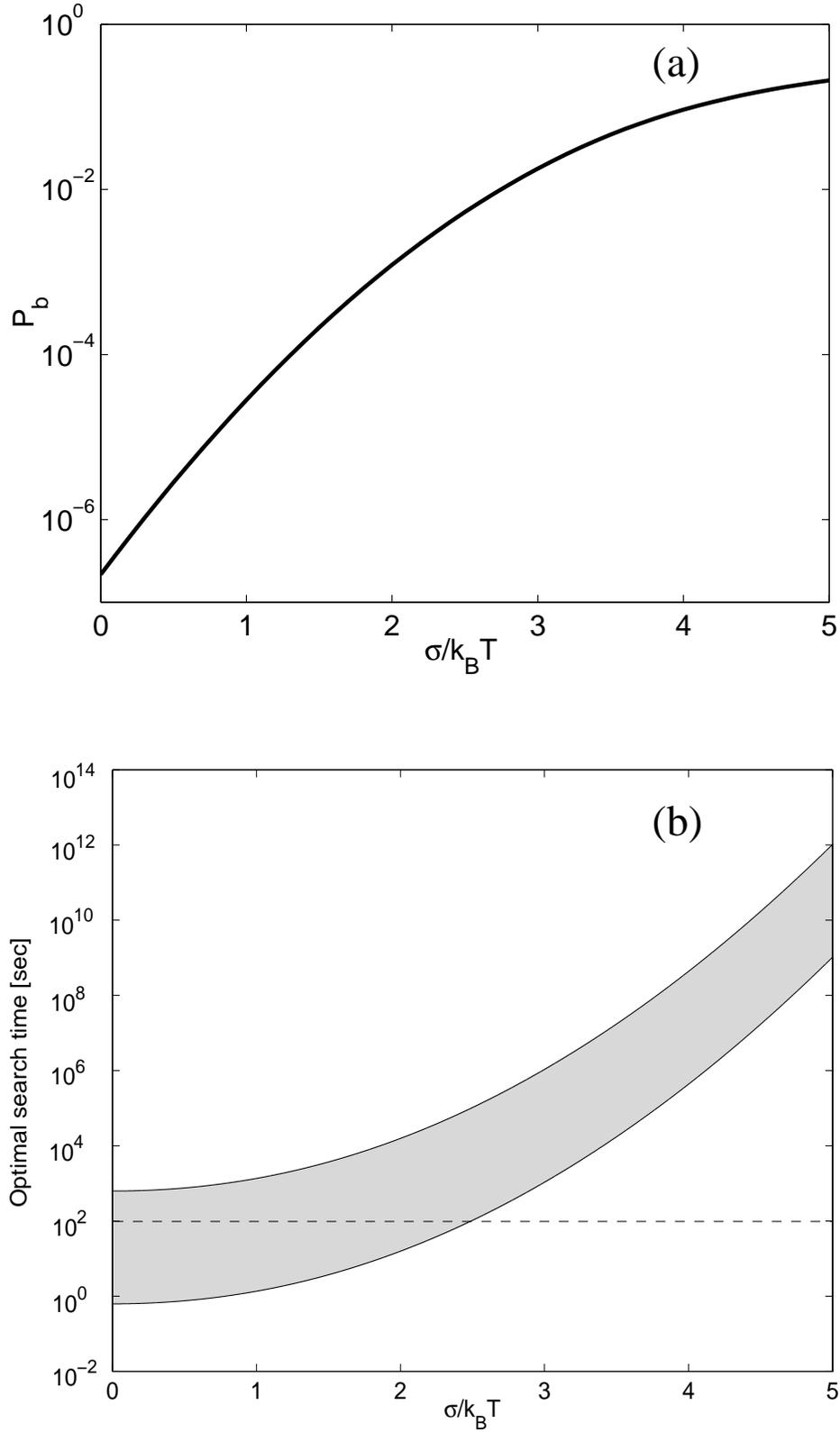}&
\end{tabular}
\caption{(a) Stability on the protein-DNA complex on the cognate site measured as the
fraction of time in the bound state at equilibrium. (b) Optimal search time as a function of the binding profile roughness, for the range of parameters $10^{-4}{\rm sec} \leq
\tau_{3d} \leq 10^{-2}{\rm sec}$, $10^{-10}{\rm sec} \leq \tau_{0} \leq
10^{-6}{\rm sec}$.} \label{fig:paradox}
\end{figure*}

In Fig.~\ref{fig:paradox}a, we present the equilibrium probability
$P_b$ of binding the strongest target site with energy $U_{\rm
min}=U_0$ \cite{hwa} as a function of $\sigma/k_BT$. In equilibrium,
$P_b$ equals the fraction of time the protein spends at the target
site:
\begin{equation}
P_b = \frac{\exp\left[-\beta U_{\rm 0}\right]}{\sum_{i=0}^M
\exp\left[-\beta U_i\right]}.
\end{equation}
Since the target site is not separated from the rest of the
distribution by a significant energy gap, $P_b$ is comparable to 1
(which is the natural requirement for a good regulatory site) only
at $\sigma$ {\em much greater than $k_BT$}.

Figure \ref{fig:paradox}b shows the optimal search time at the
corresponding values of $\sigma/k_BT$. High roughness of $\sigma
>> k_BT$ required for stability of the protein-DNA complex leads to
astronomically large search times. In contrast, a protein can
effectively search the target site {\em at $\sigma$ smaller than
$1-2 k_BT$}.

This brings us to the central result that {\em the ability to
translocate rapidly along the DNA clearly cannot comply with the
stability requirement}.

Requirement of high stability at the target site $P_b \sim 1$ (or
$P_b \sim 1/N_p$ if $N_p$ copies of the protein are present)
yields an estimate for the minimal $\sigma$,
\begin{equation}
\sigma \sim k_BT\sqrt{2\ln M} \simeq 5~k_BT,
\end{equation}
given a genome size $M = 10^6$.

From the above analysis, an obvious conflict arises: {\sl the same
energy landscape cannot allow for both rapid translocation and high
stability of states formed at sites with the lowest energy}. This
conflict is similar to the speed-stability paradox of protein folding
formulated by Gutin $et. al.$ \cite{gutin}: rapid search in
conformation space requires a smooth energy landscape, but then the
native state is unstable. In protein folding, this conflict is
resolved by the presence of a large energy gap between the native
state and the rest of the conformations
\cite{fink_book,pande_grosberg_tanaka}.

As evident from Fig.~\ref{fig:no_gap}, no such energy gap separates cognate
sites from the bulk of other (random) sites. In fact, the energy
function in the form of (\ref{eq:U}) cannot, in principle, provide a
significant energy gap. Increasing the number of TFs cannot resolve
the paradox either (see Appendix D,E). An alternative solution
must be sought.

\section{The two-mode model}
The ``search speed - stability'' paradox has already been
qualitatively anticipated by Winter, Berg and von Hippel
\cite{bvh_kin3}, who therefore concluded that a conformational
change of some sort should exist that would allow fast switching
between ``specific'' and ``non-specific'' modes of binding.  In
the non-specific mode, the protein is ``sliding'' over an
essentially equipotential surface (in our terms, $\sigma_{\rm
non-spec} = 0$) whereas site-binding takes place in the
``specific'' mode ($\sigma_{\rm spec} \gg k_BT$). A protein in the
non-specific binding mode is ``unaware'' of the DNA sequence it is
bound to.  Thus, it should permanently alternate between the
binding modes, probing the underlying sites for specificity.

This model naturally rises a question about the nature of the
conformation change. Originally, it was described as a ``microscopic''
binding of the protein to the DNA accompanied by water and ion
extrusion. However, numerous calorimetry measurements and calculations
\cite{spolar} show that such a transition is usually accompanied by a
large heat capacity change $\Delta C$. This $\Delta C$ cannot be
accounted for, unless additional degrees of freedom, namely, protein
folding, are taken into account. On-site folding of the transcription
factor may involve significant structural change
\cite{bruinsma,bruinsma1,kalodimos04} and take a time of $\sim 10^{-4}..10^{-6}~$sec 
\cite{akke} 
(compared to a characteristic on-site time of $\tau_0\sim 10^{-7}..10^{-8}~$sec). 
We conclude that conformational transition between the
two modes involves (but not limited to) partial folding of the TF.

If the TF is to probe every site for specificity in this fashion,
it would take hours to locate the native site. We note, however,
that if there was a way to probe only a very limited set of
sites, i.~e. only those having high potential for specificity, the
search time would be dramatically reduced. From the previous
section it is clear that a relatively weak site-specific
interaction (i.e. smooth landscape, $\sigma\sim k_BT$) does not
affect significantly the diffusive properties of the DNA and the
total search time. If this landscape, however, is correlated with
the actual specific binding energy landscape (with $\sigma\sim
5-6~k_BT$), the specific sites will be the strongest ones in both
modes. The protein conformational changes should occur therefore
mainly at these sites, which constitute ``traps'' in the smooth
landscape. Since such sites constitute a very small fraction of
the total number of sites, the transitions between the modes are
very rare.

\begin{figure*}[h]
\begin{center}
\includegraphics[width = 0.7\textwidth]{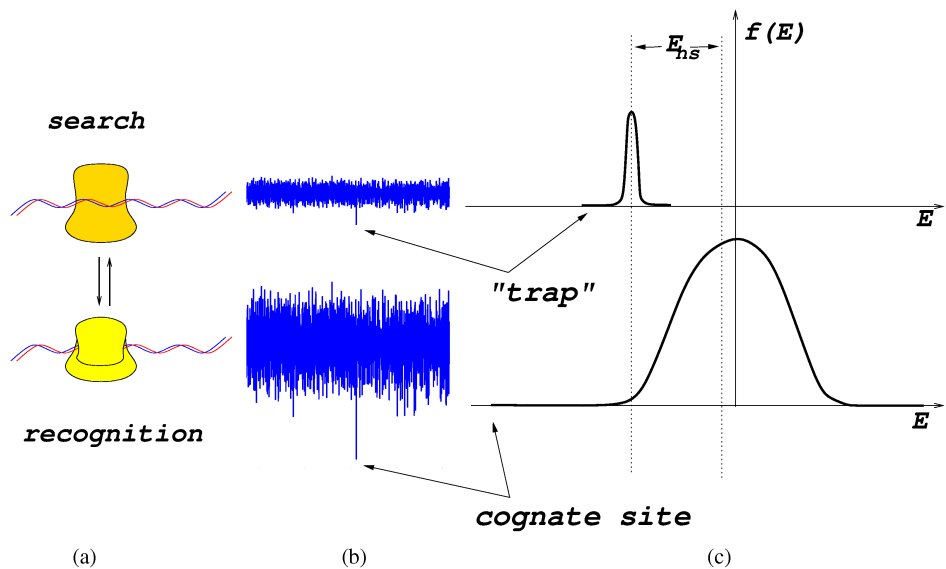}
\end{center}
\caption{Cartoon demonstrating the two-mode search-and-fold
mechanism. Top: search mode, bottom: recognition mode (a) two
conformations of the protein bound to DNA: partially unfolded (top)
and fully folded (bottom). (b) The binding energy landscape
experienced by the protein in the corresponding conformations. (c) The
spectrum of the binding energy determining stability of the protein in
the corresponding conformations.} \label{fig:mech}
\end{figure*}

We therefore suggest that there are two modes of protein-DNA
binding: the {\sl search} mode and the {\sl recognition} mode
(Fig.~\ref{fig:mech}). In the search mode, the protein
conformation is such that it allows only a relatively weak
site-specific interaction ($\sigma_{\rm s} \sim 1.0-2.0~k_BT$)
(Fig \ref{fig:mech} top).  In the recognition mode, the protein
is in its final conformation and interacts very strongly
($\sigma_{\rm r}
\geq 5~k_BT$) with the DNA (Fig \ref{fig:mech} bottom). 
If two energy profiles are strongly correlated then the lowest lying
energy levels (``traps'') in the search mode ($\leq -5~k_BT$) are likely
to correspond to the strongest sites in the recognition mode,
putatively, the cognate sites. The transitions between the two modes
happen mainly when the protein is trapped at a low-energy site of the
search landscape. In this fashion, the 1D diffusion coefficient
$D_{1d}$ is about 10--100 times smaller than the ideal limit, but the
search time in the optimal regime is reduced only by a factor of
$\sim3-10$ (see Eq.~(\ref{eq:ts_fin_min})).

The coupling between the conformational change and association at a
site with a low-energy trap is likely to take place through time
conditioning. Namely, {\sl the folding (or a similar conformation
transition) occurs only if the protein spends some minimal amount of
time bound to a certain site}. This statement is basically equivalent
to saying that the free energy barrier that the protein must overcome
to transform to the final state must be comparable to the
characteristic energy difference that controls hopping to the
neighboring sites.

The protein conformation in recognition mode should be stabilized
by additional protein-DNA interactions. If these interactions are
unfavorable, the folded structure is destabilized, then the search
conformation is rapidly restored and the diffusion proceeds as
before. If the new interactions are favorable, the folded
structure is stable and the protein is trapped at the site for a
very long time.

For this mechanism to work, transition between the two modes of search
has to be associated with significant change in the free energy
($\sim 5..10 k_BT$) of the protein-DNA complex (see Fig
\ref{fig:mech}(c)).  Such energy difference between the two states is
required to make most of the high-energy sites in the recognition mode
less favorable than in the search mode. So a protein would rather
(partially) unfold that bind an unfavorable site. As a result sites
that lay higher in energy than a certain cutoff exhibit similar
non-specific binding energy (i.e. switch into search mode of binding).
Folding of partially disorder protein loops or helices can provide
required free energy difference between the two modes. 

Efficiency of proposed search-and-fold mechanism depends on the energy
difference between the two modes, correlation between the energy
profiles and the barrier between the two states. The barrier
determines the rate of partial folding-unfolding transition. If the
barrier is too low, then the protein equilibrates while on a single
site having no effect on search kinetics. On the contrary, too high
barrier can lead to rear folding events and the cognate site can be
missed. It can be shown that proper size of the barrier provides
efficient search and stable protein-DNA complex. Alternatively, the
cognate site can lower the barrier by stabilizing the transition state
(i.e. the folding nucleus \cite{Abkevich94,Mirny01}) acting as a
catalyst of partial folding. Quantitative analysis of these factors is
beyond the scope of this study and will be published elsewhere.

\section{Discussion}

\subsection{Specificity ``for free'': kinetics vs thermodynamics.} 
The proposed mechanism of specific site location is akin to
kinetic proofreading \cite{hopf}, which is a very general concept
for a broad class of high-specificity biochemical reactions. The
required specificity is achieved in kinetic proofreading through
formation of an intermediate metastable complex that paves the way
for irreversible enzymatic reaction. If the reaction is much
slower than the life-time of the complex, then substrates that
spend enough time in the complex are subject to the enzymatic
reaction, while substrates that form short-lived complexes are
released back to the solvent before the reaction takes place. In
other words, the substrates are selected by kinetic partitioning.

In contrast to kinetic proofreading that increases equilibrium
specificity for the price of energy consumption, the
search-and-fold model doesn't require any additional source of
energy. The two mode search-and-fold model provides faster
``on-rate'' of binding while keeping the equilibrium binding constant
unchanged. Naturally, the ``off-rate'' is increased as well. This
makes our two-mode model thermodynamically ``neutral''.

\subsection{Coupling of folding and binding in molecular recognition} 
Several DNA- and ligand-binding proteins are known to
have partially unfolded (disordered) structures in the unbound
state. The unstructured regions fold upon binding to the target.
Does binding-induced folding provide any biological advantage?

The idea of coupling between local folding and site binding has
been around for some time and was recently reassessed in the much
broader context of intrinsically unstructured proteins
\cite{wright1,wright2,uversky}. Induced folding of these proteins
can have several biological advantages. First, flexible
unstructured domains have intrinsic plasticity allowing them to
accommodate targets/ligands of various size and shape. Second,
free energy of binding is required for compensation for entropic
cost of ordering of the unstructured region. A poor ligand that
doesn't provide enough binding free energy cannot induce folding
and, hence, can not form a stable complex. Williams et.al. have
suggested that unstructured domains can be result of evolutionary
selection that acts on the bound (structured) conformation, while
ignoring the unbound (unstructured) conformation \cite{Williams01}. 
Partial unfolding can also increase protein's radius of gyration and, 
hence, increase the binding rate \cite{flycast,Onuchic_binding} 

Here we propose a mechanism that suggests role of induced folding
in providing rapid and specific binding. Induced folding (or other
sort of two-state conformational transition) allows a protein to search and
recognize DNA in two different conformations providing rapid binding
to the target site. Importantly, this mechanism reconciles rapid
search for the target site with stable bound complex (see above). The
rate of induced folding can also play a role in determining the
specificity of recognition (Slutsky and Mirny, in preparation).

Structural and thermodynamic data argue in favor of distinct
protein conformations for search along non-cognate DNA and for
recognition of the target site. Proteins such as $\lambda$ cI, Eco
RV and GCN4 apparently do not fold their unstructured regions
while bound to non-cognate DNA \cite{Winkler93,Clarke91,ONeil90},
supporting our hypothesis.

Heat capacity measurements on a vast variety of protein-DNA complexes
report a large negative heat capacity change in site-specific
recognition, which is a clear indication of a phase transition. These
measurements supplemented by X-ray crystallography and NMR structural
data were interpreted by Spolar $et.~al.$ \cite{spolar} mainly in
terms of hydrophobic and conformational contributions to
entropy. Thus, folding - binding coupling is now considered a well
established effect for a large set of transcription factors.

However, real-time kinetic measurements were not performed until
recently, so that the question of the actual mechanism was left open.
Serious advances in this direction were made by Kalodimos $et.~al.$
\cite{kalodimos2,kalodimos1,kalodimos04}, 
who observed a two-step site recognition by dimeric $Lac$ repressor.
The H/D-exchange NMR data unambiguously demonstrates site
pre-selection by $\alpha$-helices bound in the major groove followed
by folding of hinge helices that bind to the minor groove elements and
complete the specific site recognition. Though the experiments in this
field were performed with a single model system, their implications
are likely to have a general character.

It should be mentioned, that no transition of this kind
is observed when the protein is unbound from DNA. A possible
reason for this can be a significant reduction of the free energy
barrier for folding, entropic in essence, that accompanies
protein-DNA association. Entropy barrier reduction is a natural
consequence of relative anchoring of the various parts of the
protein on the DNA ``scaffold''. Thermal fluctuations that the
associated protein is subject to are generally of the order of
$\sim k_BT$, and their main effect is protein translocation along
the DNA. From the above analysis, it follows that the
translocation actually takes place only if the protein encounters
barriers of $\sigma_{\rm s}\sim k_BT$ on its way.  In a
large enough collection of sites ($M \gg 1$), however, potential ``wells''
of depth $\sim\sigma_{\rm s}\sqrt{2\ln M}$ will be present. If the
``well'' depth is larger than the folding barrier height, the
probability of on-site (``in-well'') folding increases, leading
eventually to a stable complex formation. More detailed
computational analysis of coupling between folding and binding
will be published elsewhere.

\subsection{Biological implications}
Mechanism of 3D/1D search described above has several biological
implications.  Needless to say that studied model (as any quantitative
model) is a gross simplification of protein-DNA recognition {\it in
vivo}. In spite of this simplification, proposed mechanism can
be generalized to describe {\it in vivo} binding.  Here we
briefly discuss some biological implications of our model.

\subsubsection{Simultaneous search by several proteins}
If several TFs are searching for its site on the DNA, the total search
time is given by equation (\ref{eq:tsearchTF}) and is obviously
shorter than the time for a single TF. For example, if $100$ copies of
a TF are searching in parallel for the cognate site, then assuming
$k^{\rm cytoplasm}_{on} \approx 10^{8}$M$^{-1}$s$^{-1}$ and a cell of
$1\mu{\rm m}^3$ volume, we obtain the search time of $t_{s} \approx
0.1 {\rm sec}$. Increasing the number of TF molecules can further
decrease the search time, but can have harmful effects due to
molecular crowding in the cell.  Note, however, that increasing the
number of TF molecules to $100-1000$ per cell cannot resolve the
speed-stability paradox (see Fig. 5).

\subsubsection{Search inside a cell: molecular crowding on DNA and chromatin}
Above we assumed that a TF is free to slide along the DNA. {\it In
vivo} picture is complicated by other proteins and protein complexes
(nucleosomes, polymerases etc) bound to DNA, preventing a TF to slide
freely along DNA. What are the effects of such molecular crowding on
the search time?

Our model suggests that molecular crowding on DNA can have little effect
on the search time if certain conditions are satisfied. Obviously, the
the cognate shall not be screened by other DNA-bound
molecules/nucleosomes.  DNA-bound molecules can interfere with the
search process by shortening regions of DNA scanned on each round of
1D diffusion. If, however, the distance between DNA-bound
molecules/nucleosomes in the vicinity of the cognate site is greater
than $\bar{n}_{\rm opt} \sim 300 - 500 ~{\rm bp}$ (eq. (\ref{eq:nopt})
and \cite{kim}), then obstacles on the DNA do not shorten the rounds
of 1D diffusion and, hence, do not slow down the search process. Our
analysis also suggests that sequestration of part of genomic DNA by
nucleosomes can even speed up the search process.

If DNA-bound proteins are separated by more than $300-500$bp {\it
E.coli} genomic DNA can accommodate $4.6
\times 10^6 {\rm bp} / 300 {\rm bp} \approx 1.5 \times 10^4 $ 
proteins. In other words all $150$ known and predicted {\it E.coli} TF
can be simultaneously present in $100$ copies of each and search for
their cognate sites without affecting each other.  (In fact they can
be present in $200$ copies each since optimal search requires $50\%$
of proteins to be in solution at any time). On the other hand, a short
$\sim 50 {\rm bp}$ linker between nucleosomes in eukaryotic chromatin
can increase the search time about $10$-fold. Details of this
analysis will be published elsewhere.

\subsubsection{``Funnels'', local organization of sites}
Several known bacterial and eukaryotic sites tend to cluster
together. One may suggest that such clustering or other local
arrangement of sites can create a ``funnel'' in the binding energy
landscape leading to a more rapid binding of cognate sites. Our model
suggests that even if such ``funnels'' do exist, they would not
significantly speed up the search process. Proposed search mechanism 
involves $\sim M/\bar{n}_{\rm opt} \sim 10^4$ rounds of 1D/3D
diffusions. So a TF spends all the search time far from the cognate
site. Only the last round (out of $10^4$) will be sped up by the
``funnel'', leading to no significant decrease of the search time.

Local organization of sites and other sequence-dependent properties of
the DNA structure (flexibility of AT-rich regions, DNA
curvature on poly-A tracks etc) may influence preferred localization of
TFs and lead to faster on-/off- binding rates and fast equilibration on
neighboring sites (see \cite{skm_condmat} for details).

\subsubsection{Protein hopping: intersegment transfer}
Our model assumed that rounds of 1D diffusion are separated by periods
of 3D diffusion. Intersegment transfer is another mechanism that can
be involved. If two segments of DNA come close to each other, a TF
sliding along one segment can ``hop'' to another. The benefit of this
mechanism is that it significantly shortens the transfer time
$\tau_{3D}$. Several experimental evidences suggest that tetrameric
LacI, which has two DNA-binding sites, travels along DNA through 1D
diffusion and intersegment transfer.

We did not consider this mechanism because of the two following
considerations. First, it is unclear whether TFs that have only one
binding site can perform intersegment transfer. Second, for this
mechanism to work, distant segments of DNA need to come close to
each other. While DNA packed into a cell/nuclear volume crosses itself
every $\sim 500$bp, DNA in solution (at {\it in
vitro} concentrations) is unlikely to have any such self-crossings.
Hence intersegment transfer cannot explain ``faster than diffusion''
binding rates observed {\it in vitro}. This mechanism however may play
a role {\it in vivo}, especially for proteins that have multiple
DNA-binding sites.

\subsection{Proposed experiments}
Our results propose several experimentally testable predictions.

First, we predict that the maximal rate of binding is achieved when
the protein spends half of the time in solution and half sliding along
the DNA. This result can be readily verified experimentally by
measuring concentration of free protein in solution that contains DNA
but no cognate site. We also show how the search time depends on the
energy of non-specific binding, that, in turn, can be controlled by
ionic strength of solution or by engineering proteins with stronger or
weaker non-specific binding. {\it In vivo} observation of the "50/50"
rule would suggest that proteins are optimized by evolution for rapid
search.

Second, we show how binding rate depends on the average travel time
between two random segments of DNA, $\tau_{3d}$. Time $\tau_{3d}$
depends on the DNA concentration and domain organization of DNA. By
changing DNA concentration and/or DNA stretching in a single molecule
experiment one can alter $\tau_{3d}$ and thus study the role of DNA
packing on the rate of binding. This effect has implications for DNA
recognition {\it in vivo}, where DNA is organized and
domains. Similarly one can experimentally measure and compare with
analytical predictions the binding rate in the presence of other
DNA-binding proteins or nucleosomes.

Single molecule experiments and AFM/SFM imaging allow direct
observation of protein trajectory and measurement of the 1D
diffusion coefficient, $D_{1d}$ on non-cognate DNA. Our formalism,
in turn, allows to calculate the spectrum of specific binding
energy given $D_{1d}$. Such measurements can be direct tests of
our conjecture that 1D search along non-cognate DNA proceeds along
a ``smoother'' energy profile.

Third, using protein engineering one can stabilize unstructured
regions of DNA-binding proteins (e.g. $\lambda$ cI, Eco RV and
GCN4) and study binding rates of these engineered rigid protein.
Such experiments can test proposed search-and-fold mechanism and
shed light on the role of unstructured regions in determining
stability, specificity and binding rates.

We also suggest that proteins bound to non-cognate DNA are not fully
ordered. Unfortunately very few studies
\cite{kalodimos2,kalodimos1,kalodimos04} have addressed the
mechanisms of binding to non-cognate DNA. More studies of structures,
thermodynamics and dynamics of proteins bound to non-cognate DNA will
deepen our understanding of specific protein-DNA recognition.

\section{Conclusions} 
We have developed a quantitative model of protein-DNA interaction that
provides an insight into the mechanism of fast target site
location. We found the range of parameters (specific and non-specific
binding energies) that are crucial for fast search and, hence, robust
functioning of gene transcription. Paradoxically, realistic energy
cannot provide both rapid search and strong binding of a rigid
protein. This allowed us to formulate speed-stability paradox of
protein-DNA recognition (which is similar to famous Levinthal paradox
of protein folding). To resolve this paradox we proposed a
search-and-fold mechanism that involves the coupling of protein
binding and protein folding.

Proposed mechanism has several important biological implications
explaining how a protein can find its site on DNA {\it in vivo} in the
presence of other proteins and nucleosomes, and by simultaneous search
of several proteins. Our model provides, for the first time,
quantitative framework for analysis of kinetics of transcription factor
binding and, hence, gene expression. Importantly, our model links
molecular properties of transcription factors to the timing of
transcription activation. Proper understanding of the entire mechanism
will hardly be possible without further experimental effort in these
directions.

\acknowledgments
We are thankful to A. Finkelstein, M. Kardar, W. Bialek and A. van
Oijen for useful discussions. LM is an Alfred P. Sloan Research
Fellow.

\section*{Appendix A: Diffusive properties of the DNA.}\label{app:diff}

The  derivation consists of two steps.  First,  we describe the
random walk along the DNA in terms of number of steps. Next, we
calculate the mean   time between successive steps  in  a random
energetic landscape which provides  the time - scale  for the
problem. Such a decoupling, strictly speaking, does not hold when
the number of steps is small, i. e. when the number of visited
sites is small and the random quantities are not averaged
properly.  However, since we  are dealing with large numbers of
steps ($\sim 10^5 - 10^6$) this approach is legal, which is also
confirmed by numerical simulations.

\subsection*{The MFPT.}

To derive the diffusion law, we calculate the mean first passage
time (MFPT) from site $\#0$ to site $\#L$, defined as the mean
number of steps the particle is to make in order to reach the site
$\#L$ {\sl for the first time}. The derivation here follows the
one in \cite{kehr}.

Let $P_{i,j}\left(n\right)$ denote the   probability to start at
site $\#i$  and  reach  the site $\#j$ in   exactly $n$ steps.
Then, for example,
\begin{equation}\label{eq:rec_1}
P_{i,i+1}\left(n\right) = p_i T_i\left(n - 1\right),
\end{equation}
where  $T_i\left(n\right)$ is defined as  the probability of
returning to the $i$-th site after $n$ steps {\sl without}
stepping to the right of it. Now,  all the  paths contributing  to
$T_i\left(n -  1\right)$ should start with the  step to the left
and then reach the site $\#i$ in $n-2$ steps,  not  necessarily
for   the first time.   Thus,   the probability $T_i\left(n -
1\right)$ can be written as
\begin{equation}\label{eq:rec_2}
T_i\left(n       -      1\right)     =       q_i       \sum
_{m,l} P_{i-1,i}\left(m\right)T_i\left(l\right)\delta_{m+l, n-2}.
\end{equation}
We now introduce generating functions
\begin{equation}
\tilde P_{i, j}\left(z\right) = \sum_{n=0}^{\infty}
z^n~P_{i,j}\left(n\right), \qquad \tilde T_{i}\left(z\right) =
\sum_{n=0}^{\infty} z^n~T_{i}\left(n\right).
\end{equation}
One can easily show (see e.~g.~\cite{goldhirsh}) that
\begin{equation}
\tilde P_{0, L}\left(z\right) = \prod_{i = 0}^{L-1} \tilde P_{i,
i+1}\left(z\right).
\end{equation}
Knowing $\tilde  P_{i,  i+1}\left(z\right)$, one calculates   the
MFPT straightforwardly as
\begin{eqnarray}
\bar{t}_{0,L} = \frac{\sum_n n P_{0, L}\left(n\right)}{\sum_n
P_{0, L}\left(n\right)} = \left[\frac{d}{dz}\ln \tilde P_{0,
L}\left(z\right) \right]_{z = 1} \nonumber \\
= \sum_{i = 0}^{L-1}\left[\frac{d}{dz}\ln \tilde P_{i,
i+1}\left(z\right) \right]_{z = 1}.
\end{eqnarray}
Using (\ref{eq:rec_1}) and  (\ref{eq:rec_2}), we obtain  the
following recursion relation for $\tilde P_{i,
i+1}\left(z\right)$:
\begin{equation}\label{eq:P_i}
\tilde P_{i, i+1}\left(z\right) = \frac{zp_i}{1 - zq_i \tilde
P_{i-1, i}\left(z\right) }.
\end{equation}
To solve      for   $\bar{t}_{0,L}$,  we  must    introduce
boundary conditions. Let $p_0 = 1,~q_0 = 0$, which is equivalent
to introducing a  reflecting wall  at  $i  = 0$.    This  boundary
condition  clearly influences the solution   for short times  and
distances.  However, as numerical   simulations  and   general
considerations  suggest,  its influence relaxes quite fast, so
that for  longer times, the result is clearly independent of the
boundary. The  benefit of setting $p_0 = 1$ becomes clear when we
observe that
\begin{equation}
\tilde P_{0, 1}\left(1\right) = 1 \qquad \Rightarrow \qquad
\forall~i\qquad \tilde P_{i, i+1}\left(1\right) = 1.
\end{equation}
Hence,
\begin{equation}
\bar{t}_{0,L} = \sum_{i = 0}^{L-1} \tilde P_{i,
i+1}'\left(1\right).
\end{equation}
The recursion relation   for $P_{i,  i+1}'\left(1\right)$  is
readily obtained from (\ref{eq:P_i}) :
\begin{equation}
\tilde P_{i, i+1}'\left(1\right) = \frac{1}{p_i} + \frac{q_i}{p_i}
\tilde P_{i-1, i}'\left(1\right) = 1 + \alpha_i \left[1 + \tilde
P_{i-1, i}'\left(1\right)\right],
\end{equation}
with  $\alpha_i     \equiv   p_i/q_i$.  Thus,   the    expression
for $\bar{t}_{0,L}$ is obtained in closed form
\begin{equation}\label{eq:mfpt_1}
\bar{t}_{0,L} = L + \sum _{k = 0}^{L - 1} \alpha_k + \sum _{k =
0}^{L - 2} \sum _{i = k + 1}^{L - 1} \left( 1 + \alpha_k \right)
\prod_{j = k + 1}^{i} \alpha_j.
\end{equation}
This solution expression gives MFPT in terms of a given
realization of disorder producing a certain set of probabilities
$\{p_i\}$, whereas we are interested in the behavior averaged
over all realizations of disorder.  The cumulative products in
(\ref{eq:mfpt_1}) reduce to the two form $e^{\beta\left(U_{i} -
U_{j}\right)}$, which after being averaged over {\em
uncorrelated} Gaussian disorder produce a factor of
$e^{\beta^2\sigma^2}$.
After the summations are carried out, the expression for MFPT
becomes for $L \gg 1$
\begin{equation}\label{eq:mftp_2}
\left\langle\bar{t}_{0,L}\right\rangle \simeq L^2
e^{\beta^2\sigma^2}.
\end{equation}
Thus,  the diffusion  law appears  to  be  the  classical one,
with  a renormalized diffusion coefficient.

\subsection*{The time constant.}

Consider a particle at site $\#i$. The particle will eventually
escape to one of the neighboring sites $\#(i\pm 1)$, the escape
rate being 
\begin{equation}
r_i = \omega_{i,i+1} +  \omega_{i,i-1}.
\end{equation}
To calculate the characteristic diffusion time constant $\langle
\tau \rangle$, this rate should be averaged over all
configurations of disorder $\{U_i\}$.  To obtain an analytic
expression for the $\langle \tau \rangle$, we assume the form
\begin{equation} 
\omega_{i,i\pm 1} = \nu e^{ -\beta\left(U_{i\pm 1} -U_{i}\right)}
\end{equation}
{\em for both $U_{i\pm 1} > U_{i}$ and $U_{i\pm 1} < U_{i}$ }, as
opposed to the form (\ref{eq:prob}). Numerics show that this
approximation introduces an up to $\sim15\%$ error for small
values of $\beta\sigma$ and is practically exact for
$\beta\sigma>2$.  Thus,
\begin{equation}
r_i =  \frac{1}{2\tau_0}\left( e^{ -\beta\left(U_{i+1} -
U_{i}\right)} + e^{ -\beta\left(U_{i-1} - U_{i}\right)}\right),
\end{equation}
where $\tau_0 = 1/(2\nu)$.  The mean time between the successive
steps can be calculated therefore as the average over all
possible configurations of $U_i$, $U_{i \pm 1}$ of the reciprocal
of the escape rate, i.~e.
\begin{equation}
\left\langle \tau \right\rangle = \left\langle \frac{1}{r_i}
\right\rangle =
2\tau_0\int_{-\infty}^{\infty}dU_idU_{i+1}dU_{i-1}\frac{f\left(U_i\right)f\left(U_{i+1}\right)f\left(U_{i-1}\right)}{e^{-\beta\left(U_{i+1}-U_{i}\right)}+
e^{-\beta\left(U_{i-1} - U_{i}\right)}}.
\end{equation}
Assuming as  above  Gaussian  energy   statistics, this  integral
is evaluated as follows
\begin{equation}
\left\langle \tau \right\rangle =
\frac{\tau_0~e^{\beta^2\sigma^2/2}}{\pi}\int_{-\infty}^{\infty}dxdy\frac{e^{-(x^2+y^2)/2}}{e^{-\beta\sigma
x}+ e^{-\beta\sigma y}}.
\end{equation}
After the change of variables
\begin{equation}
s = \frac{1}{\sqrt{2}}(x+y), \qquad t = \frac{1}{\sqrt{2}}(x-y),
\end{equation}
the integral factorizes leading to
\begin{eqnarray}
\left\langle \tau \right\rangle &=&
\frac{\tau_0~e^{\beta^2\sigma^2/2}}{2\pi}
\int_{-\infty}^{\infty}ds~e^{-s^2/2 + \beta\sigma s/\sqrt{2}}
\int_{-\infty}^{\infty}dt\frac{e^{-t^2/2}}{\cosh (\beta\sigma
t/\sqrt{2})}
\nonumber \\
 &=&
 \frac{\tau_0~e^{3\beta^2\sigma^2/4}}{\sqrt{2\pi}}\int_{-\infty}^{\infty}dt~e^{-t^2/2
 - \ln\left[\cosh (\beta\sigma t/\sqrt{2})\right]}
 \\ &\simeq& \frac{\tau_0~e^{3\beta^2\sigma^2/4}}{\sqrt{2\pi}}
\int_{-\infty}^{\infty}dt~e^{-t^2(1 + \beta^2\sigma^2/2)/2}
 =            {\tau_0}~e^{3\beta^2\sigma^2/4}\left[1                 +
 \beta^2\sigma^2/2\right]^{-1/2} \nonumber
\end{eqnarray}

Now,      multiplying   (\ref{eq:mftp_2})     by  $\left\langle
\tau \right\rangle$, we obtain the diffusion coefficient as
\begin{equation}
D_{1d}\left(\sigma\right) \simeq \frac{1}{2\tau_0}\left(1 +
\frac{\beta^2\sigma^2}{2}\right)^{1/2} e^{- 7\beta^2\sigma^2/4}.
\end{equation}

\section*{Appendix B: Non-specific energy}
To find how the non-specific energy $E_{\rm ns}$ is related to the
average time $\tau_{1d}$ protein spends scanning a single region
of the DNA we use simple observation that
\begin{equation}\label{eq:E_ns_text}
\left\langle\sum_i \tau_i r_i \right\rangle = 1 \ \ \ \
\left\langle \sum_i \tau_i\right\rangle = \tau_{1d}
\end{equation}
which states that "on average" protein dissociates ones from the
region it scans.

Since some massive hopping from site to site  takes  place before
the particle eventually    dissociates,  the  dissociation rates
and, consequently,  the   non-specific binding  energy should
satisfy the following equation
\begin{equation}\label{eq:E_ns}
\left\langle\sum_i \tau_i r_i \right\rangle =
\frac{1}{\tau_0}\left\langle\sum_i \tau_ie^{ -\beta\left(E_{\rm
ns} - U_i\right)}\right\rangle =
\frac{1}{\tau_0}\int_{-\infty}^{\infty} e^{ -\beta\left(E_{\rm ns}
- U\right)} \tau \left(U\right) f \left(U\right) dU = 1,
\end{equation}
and this subject to a condition
\begin{equation}\label{eq:tau_i}
\left\langle \sum_i \tau_i\right\rangle = \int_{-\infty}^{\infty}
\tau \left(U\right) f \left(U\right) dU = \tau_{1d}.
\end{equation}
Where $\tau_i$ is the time TF spends at the $i$-th site and
$\tau_{1d}$ is the average time of 1D search to dissociation.The
average lifetime $\tau_i = \tau \left(U_i\right)$ at that site is
proportional to $\exp\left(-\beta U_i\right)$.  In this specific case,
the particle usually escapes to one of the neighboring sites, and we
should average over their energies.  Hence, the explicit
form $\tau \left(U\right)$ as calculated from (\ref{eq:tau_i}) is
\begin{equation}
\tau \left(U\right) = \tau_{1d}e^{-\beta^2\sigma^2/2}e^{-\beta U}.
\end{equation}
Substituting this into (\ref{eq:E_ns}), we have
\begin{equation}
 \frac{\tau_{1d}}{\tau_0}e^{- \frac{1}{2}\beta^2\sigma^2 - \beta E_{\rm ns}} = 1,
\end{equation}
or
\begin{equation}\label{eqA:E_ns_fin}
E_{\rm ns} = k_BT\left[\ln\left(\frac{\tau_{1d}}{\tau_0}\right) -
\frac{1}{2}\left(\frac{\sigma}{k_BT}\right)^2\right].
\end{equation}

Next we recall that, in the optimal regime, $\tau_{1d} =
\bar{\tau}_{3d}$.  Thus, to ensure optimal performance, $E_{\rm
ns}$ should be equal the expression in (\ref{eqA:E_ns_fin}) with
$\tau_{1d}$ replaced by $\bar{\tau}_{3d}$:
\begin{equation}\label{eqA:E_ns_opt}
E_{\rm ns} = k_BT\left[\ln\left(\frac{\tau_{3d}}{\tau_0}\right) -
\frac{1}{2}\left(\frac{\sigma}{k_BT}\right)^2\right].
\end{equation}
The meaning of this relation is quite transparent. The logarithm
gives $E_{\rm ns}$ in   a system  with  zero  or constant specific
binding energy. The second term introduces suppression of  $E_{\rm
ns}$ due to disorder, so that the dissociation  events in  a
system with  disorder are  more  frequent  to compensate partially
for   the  1D diffusion slowdown. This  relation obviously holds
as long   as $E_{\rm  ns}  > 0$. Negative values of $E_{\rm ns} $
mean simply that the non-specific interaction became overshadowed
by the specific one  and has no direct physical sense anymore.

Since for a given value of $\sigma$, the non-specific binding
controls the dissociation rate, the search time will deviate from
the optimum if $E_{\rm ns}$ moves from this predetermined value.
In Fig.3a we plot the search time as a function of the
non-specific binding energy for different values of $\sigma$.

We now define the {\sl tolerance factor} $\zeta$ as the ratio
between the maximal acceptable value of the search time $t_s$ and
the minimal time $t_{s0}$. Experimental data suggest $\zeta \leq
5$, but we for the moment allow for much larger values of
$\zeta\sim 10 - 100$ (this can be done when, for instance, there
are many protein molecules searching in parallel). As we can see
from Fig.3a, for each value of $\sigma$, there is a range of
possible values of $E_{\rm ns}$ such that the resulting search
time is within the region of tolerance.  This range is easily
calculated  producing the values of non-specific energy between
\begin{equation}
E_{\rm ns}^{\pm}\left(\sigma, \zeta \right) =
 \frac{2}{\beta}\ln
\left[\sqrt{\frac{D_{1d}(\sigma)\bar{\tau}_{3d}}{D_{1d}(0)\tau_0}}
\left(\zeta \pm \sqrt{\zeta^2 - \frac{D_{1d}(0)}
{D_{1d}(\sigma)}}\right)\right]
 -\frac{\sigma^2\beta}{2}
\end{equation}

\section*{Appendix C: Role of DNA conformation}
Central parameter here is $\tau_{3d}$, the interval of time
between a dissociation of the protein from DNA till the next
binding to DNA. Exact calculation of $\tau_{3d}$ is a very
difficult task, considering the nontrivial packaging of the DNA
molecule inside a bacterial cell, electrostatic effects and the
inhomogeneity of the cytoplasm.
\begin{figure*}[hbt]
\begin{center}
\includegraphics[width = 0.7\textwidth]{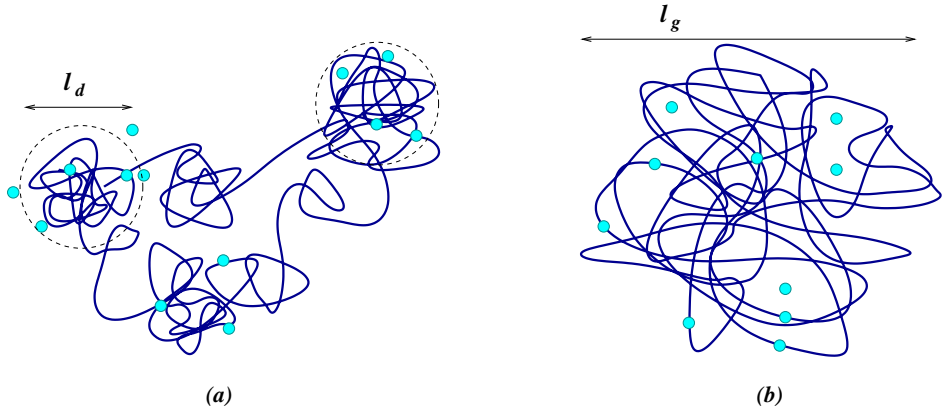}
\caption{Effect of DNA conformation on the effective diffusion
distance: (a) Single globule; (b) Multi-domain
conformation.}
\label{fig:dna_conf}
\end{center}
\end{figure*}
Considering the microscopic picture one can easily obtain a
reasonable estimate of $\tau_{3d}$ as a characteristic time of 3D
diffusion across the nucleoid (the region of a bacterial cell to
which the DNA is confined). The corresponding diffusion length
depends on the conformation of the DNA molecule. Indeed, if the
DNA molecule was a single homogeneous globule, there would be a
single relevant length scale, which is the molecule characteristic
size $l_{\rm m}$ (the gyration radius). On the other hand, as
Fig.~\ref{fig:dna_conf} shows, diffusion of a protein molecule
inside a more realistic non-homogeneous multi-domain molecule
involves at least one additional length scale $l_{\rm d}$, which
is a characteristic size of a domain. These two lengths may differ
by a factor of $\sim 10$ \cite{nucleoid}, making the ratio of the
resulting diffusion times $\tau^m_{3d}/\tau^d_{3d}\sim 10^2$. In
the original problem (a single protein molecule searching for a
single site on the DNA), the search process is dominated by the
larger time-scale, since at least few domains must be explored
before the target site is located. However, there are usually
about $10^2$ TF molecules present in a cell, so it is reasonable
to assume that the domains are scanned in parallel, making the
inter-domain transfer processes irrelevant.

\section*{Appendix D: Stability requirement}

In fact, it is not hard to estimate analytically the
$(\sigma/k_BT)$ ratio for a genome of length $M$ such that the
probability of binding to the lowest site is comparable to the
probability of binding to the rest of the genome. , i. e. their
contributions to the partition function are of the same order of
magnitude. The partition sum for the Gaussian energy level
statistics is.
\begin{eqnarray}
\Omega &=& \frac{M}{\sqrt{2\pi \sigma^2}}\int_{-\infty}^{\infty}
e^{-\beta U - U^2 /(2\sigma^2)} dU =
Me^{\beta^2\sigma^2/2}\sim \nonumber \\
&\sim& \exp\left[-\beta U_{\rm min}\right] \sim
\exp\left(\beta\sigma\sqrt{2\ln M}\right)
\end{eqnarray}
so that for $M = 10^6$
\begin{equation}
\sigma \sim k_BT\sqrt{2\ln M} \simeq 5~k_BT.
\end{equation}
Strictly speaking, for a large though finite set of energy levels,
the integration limits are cut off at $\pm\sigma\sqrt{2\ln M}$ so
that for $\beta\sigma \gg \sqrt{\ln M}$ the partition function is
dominated by the lower edge of the distribution. The estimate for
$\beta\sigma$ gives therefore the crossover value between the
regime of multiple-site contribution to $\Omega$ and the regime
with single-site domination\footnote{In the Random Energy Model
\cite{derr}, the analog of this effect would be the
thermodynamical freezing.}.

If $N_p$ proteins are searching and binding a single target site,
then the probability of being occupied is given by
\begin{equation}\label{eq:P_N_p}
    P(N_p) = 1 - \left(1-P_b\right)^{N_p} \approx N_p P_b
\end{equation}
where $P_b$ is the probability of the site being occupied by a
single protein (eq 19 of the paper) and approximation is for $P_b
<< 1/N_p$. As evident from Fig 4b, requirement of the rapid search
is satisfied if $P_b(\sigma/T\approx 1) \sim 10^{-5}$. An
unfeasible amount of $10^4$ copies of a single TF are required to
saturate such weak binding site.

\section*{Appendix E: Energy Gap}
Large energy gap between the cognate site $\vec{s}_c$ and the bulk
of genomic sites would solve the paradox of rapid search and
stability. One may seek parameters $\epsilon(j,s)$ of the energy
function
\begin{equation} \label{eqA:U}
U(\vec{s}=s_i,..s_{i+l-1}) = \sum_{j=1}^{l} \epsilon(j,s_j),
\end{equation}
to maximize the energy gap by minimizing the Z-score
\begin{equation}\label{eq:gap}
Z(\vec{s}_c)=\frac{U(\vec{s}_c)-\langle U \rangle }{\sigma},
\end{equation}
where averaging and variance is taken over all possible sequences
of length $l$ (or over genomic words of length $l$). It's easy to
see that $Z(\vec{s}_c)$ is minimal if
\begin{equation}
    \epsilon^{\rm opt}(j,s) = - \delta(s,{s_c}_j)
\end{equation}
where $\delta(x,y)$ is Kronecker delta. For $K$ types of
nucleotides assuming their equal frequency in genome we obtain the
maximal reachable energy gap of
\begin{equation}
    Z^{\rm min} = - \sqrt{l K}.
\end{equation}
For $K=4$ and $l \approx 8$ we get $Z^{\rm min} \approx - 5$. For
the genome of $10^6-10^7$bp the energy spectrum of the genomic DNA
ends at $Z \approx -5 $. While sufficient to provide stability of
the bound complex (see main text), such energy gap is unable to
resolve the search-stability paradox.

\section*{Appendix F: Diffusion in water and in cytoplasm}
The diffusion coefficient of a protein molecule in water can be
estimated as \cite{landafshitz}
\begin{equation}
D \simeq \frac{k_BT}{3\pi\eta d},
\end{equation}
where $d$ is the diameter of the molecule and $\eta$ is the water
viscosity. Setting $\eta \sim 10^{-2}~\rm{g/(sec\cdot cm)}$ and $d
\sim 10~\rm{nm}$, we obtain at room temperature
\begin{equation}
D \sim 10^2~\mu\rm{m^2/sec}.
\end{equation}

Diffusion coefficient measurements for GFP in $E.~coli$
\cite{elowitz} produce values of about $1-10~\mu\rm{m^2/sec}$.
This difference in diffusion coefficients may account for more
than order of magnitude difference in the theoretically calculated
and measured target location times.

\bibliographystyle{apsrmp}
\bibliography{search_dyn}

\end{document}